\documentclass[12pt]{article}

\begin{document}

\centerline{\Large P.Grinevich, A.Mironov, S.Novikov\footnote{All
Novikov's works quoted here, can be taken directly from his homepage
http://www.mi.ras.ru/ $\tilde{} $ snovikov, click {\bf
Publications}. Our e-mails are pgg@landau.ac.ru (Grinevich),
mironov@nsc.math.ru (Mironov), novikov@ipst.umd.edu (Novikov)}}

 \centerline{\bf 2D Schrodinger Operator, (2+1)
Systems and New Reductions.}

\centerline{\bf  The 2D Burgers Hierarchy and Inverse Problem
Data\footnote{This work was accepted for publication in the Journal
Russian Math Surveys. It will appear in 2010, v 65 n 3}}

\vspace{1cm}

{\bf Abstract}. The Theory of (2+1) Systems based on 2D Schrodinger
Operator was started by S.Manakov, B.Dubrovin, I.Krichever and
S.Novikov in 1976 (see[1,2]). The Analog of Lax Pairs introduced in
[1], has a form $L_t=[L,H]-fL$ (''The $L,H,f$-triples'') where
$L=\partial_x\partial_y+G\partial_y+S$ and $H,f$-some linear PDEs.
Their Algebro-Geometric Solutions and therefore the full higher order 
hierarchies were constructed in [2]. The Theory of 2D Inverse Spectral 
Problems for the Elliptic Operator $L$ with $x,y$ replaced by 
$z,\bar{z}$, was started in [2]:  The Inverse
Spectral Problem  Data are taken from the  complex ''Fermi-Curve''
consisting of all Bloch-Floquet Eigenfunctions  $L\psi=const$. Many
interesting systems were found later [3]. However, specific properties of 
the very first system offered in [1] for the verification of new method only, 
were not studied more than 10 years until B.Konopelchenko found in 1988
(see [6]) analogs of Backund Transformations for it. He  pointed out
on the ``Burgers-Type Reduction''\footnote{Unfortunately, in the
work [6] this system is presented as a new one. At the same time,
the work [1] where it was originally found, is included in the list
of references of [6]. The work [2] where Algebro-Geometric Solutions
of such systems were found in 1976, is not quoted at all in [6]. It
leads to the wrong impression that there was no development till
1980, reversing the chronological order of participating authors to
the opposite}. Indeed, the present authors quite recently found
very interesting extensions, reductions and applications of that
system both in the theory of nonlinear evolution systems (The
Self-Adjoint and 2D Burgers Hierarhies were invented, and
corresponding reductions of Inverse Problem Data found) and in the
Spectral Theory of Important Physical Operators (''The Purely
Magnetic 2D Pauli Operators''). We call this system GKMMN by the
names of  authors who studied it.

\vspace{1cm}

Let us consider the 2nd order operators $L,H$ and scalar function
$f$, reduced to the following form by the gauge transformations
$$L=\partial_x\partial_y+G\partial_y+S,
H=\partial_x^2+F\partial_y+A$$ Using $L,H,f$-triple (see Abstract)
we define corresponding (2+1)nonlinear evolution  system. We call it
''The GKMMN System''. Making calculation, we obtain following

{\bf Proposition.} The GKMMN System has a form(I)
$$G_t=G_{xx}-G_{yy}+(F^2)_x-(G^2)_x-A_x+2S_y,S_t=-S_{xx}+S_{yy}+2(GS)_x-2(FS)_y$$
$$F_x=2G_y,A_y=2S_x,f=2G_x-F_y$$

Let us formulate some useful corollaries of that system.

{\bf Corollary 1.}The system GKMMN is compatible with the purely
real reduction where all coefficients are real.

{\bf Corollary 2.} The system GKMMN admits a Reduction $S=0$. We
call it  {\bf ''The 2D Burgers System''} and denote by $B_2$. The
whole Hierarchy can be naturally constructed after Theorem 2 below.

{\bf Corollary 3.} For the GKMMN system and its stationary problem
the elliptic operator $H$ can be self-adjoint only in the trivial
cases reducible to the functions of one variable. Here
$H=\Delta+F\partial_y+A$ is such that the magnetic field $B=F_x/i,
i^2=-1,$ and electric potential $U=A-F^2/4-F_y/2$ are real and
smooth.

Proof of this corollary requires calculations. Under these
restrictions the system GKMMN became strongly over-determined
leading to the complete degeneration.

{\bf Conjecture.} For the smooth periodic second order self-adjoint
elliptic 2D operators  the complete complex manifold of the
Bloch-Floquet Eigenfunctions  $W$ (except some trivial cases
reducible to one variable), cannot  contain  Zariski Open Part of
the Complex Algebraic Curve $\Gamma\subset W$ except of the levels
$\epsilon=const$ found in 1976 in [2].

{\bf Corollary 4.} The substitution $$G=-(\log c)_x, F=-2(\log c)_y,
A=-2u_x,S=-u_y$$ transforms our system into the following system
(II):
$$[(c_t-c_{xx}+c_{yy})c^{-1}]_x=2(u_{yy}-u_{xx}),
u_t=u_{yy}-u_{xx}+2(u_yc_x/c)_x-2(u_yc_y/c)_y$$ The $B_2$ Reduction
$S=0$ reduces system to the linear form (III):
$$c_t-c_{xx}+c_{yy}=(U(x)+V(y))c$$ exactly in the same way as the
ordinary 1D Burgers System (i.e. our system with $U=V=0$ not
depending on the variable $y$).

The spectral meaning of this variables and substitution will be
clarified below for the Algebro-Geometric (AG) Solutions immediately
leading to the full Hierarchy of such systems.

{\bf The Algebro-Geometric (AG) Inverse Spectral Problem Data.}

Take Riemann Surface $\Gamma$ with 2 ''infinite'' points
$\infty_1,\infty_2$ and local parameters $1/k_1,1/k_2$ near them,
$1/k_j(\infty_j)=0$. Select the ''Divisor of poles'' $D=P_1+...+P_g$
in $\Gamma$. Construct ''The 2-point Baker-Akhiezer Function''
$\psi(P,x,y,t)$ invented in [2]. It should be meromorphic in the
variable $P\in \Gamma$ outside of infinities, with divisor of poles
D which is $x,y,t$-independent. Its asymptotic behavior near
infinities is following:
$$\psi=ce^{k_1x+k_1^2t}(1+v/k_1+O(1/k_1^2)),\psi=e^{k_2y+k_2^2t}(1+u/k_2+O(1/k_2^2))$$
This function satisfies to the equation $L\psi=0$ and to the
(2+1)-systems (I,II,III) with parameters $(c,u)$ entering it.

{\bf The Real AG Reduction} of (I) is following: There is an
antiholomorphic involution $\sigma:\Gamma\rightarrow\Gamma,
\sigma^2=1$, such that
$$\sigma(\infty_j)=\infty_j,\sigma^*(k_j)=-\bar{k_j},\sigma(D)=D$$
Easy to formulate conditions such that real solutions (written
through the $\Theta$-functions) are smooth nonsingular. For the
dense family of data they are periodic. In general they are
quasiperiodic as usual.

{\bf The Stationary AG Solutions} are such that $[L,H]=fL$ and
$H\psi=\lambda(P)\psi$ where $H$ is an elliptic operator as above.
They correspond to the algebraic curves $\Gamma$ with algebraic
function $\lambda$ having exactly 2 poles on $\Gamma$ of the second
order in both infinite points $\infty_1,\infty_2$. However, they are
non-self-adjoint in the nontrivial cases.

{\bf The Burgers Reduction $B_2$ is especially interesting. Here
$S=u_y=0$.}

{\bf Theorem 1.} Take reducible Riemann Surface
$\Gamma=\Gamma'\bigcup\Gamma''$ such that
$$\Gamma'\bigcap\Gamma''=Q=Q_0\bigcup...\bigcup
Q_k,\infty_1\in\Gamma'\infty_2\in\Gamma''$$  and divisor $$
D=D'+D'', D'\subset\Gamma',D''\subset\Gamma'', |D'|=g'+k,\\
|D''|=g''$$ where $g'$=genus of $\Gamma'$, $D''$=genus of
$\Gamma''$, all points $\infty_j, Q,D$ are distinct. Construct $\psi
$ as a standard one-point Baker Akhiezer function $\psi''$ on
$\Gamma''$ with divisor $D''$ and asymptotic
$\psi''=e^{k_2x+k_2^2t}(1+O(1/k_2))$ On the part $\Gamma'$ our
function $\psi$ should coincide with $\psi'$. It has the divisor of
poles $D'$, asymptotic $\psi'=ce^{k_1y+k_1^2t}(1+O(1/k_1)$ and
conditions (*)
$$\psi'(Q_s)=
\psi''(Q_{\sigma(s)})$$ where $\sigma$ is some permutation of the
set $Q$. Then we have $L(\psi)=0$ and $(L_t-[L,H])\psi=0$ with
$S=u_y=0$.

{\bf Remark.} We can drop the surface $\Gamma''$ and divisor $D''$.
Take any set of solutions $\psi''_s(x,t)$ to the equation
$\psi''_{s,t}=\psi''_{s,xx}, s=0,1,...,k$. Define $\psi'(x,y,t,P)$
using conditions $\psi'(x,y,t,Q_s)=\psi''_s(x,t)$ instead of
conditions (*). Our function $\psi=\psi'$ satisfies to the equations
$L\psi=0, L_t=[L,H]-fL$ for all points $P\in \Gamma'$, and
$S=u_y=0$.

Corresponding hierarhy with higher times we call ''The 2D Burgers
Hierarhy'' $B_2$.

 There are 2 cases in our theory:

 1.$(x,y)\in R$. This is the system  $GKMMN-I$ and reduction $B_2-I$

 2.$x\rightarrow z,y\rightarrow \bar{z}$ and $\partial_x\rightarrow \partial=\partial_x-i\partial_y,
 \partial_y\rightarrow\bar{\partial}= \partial_x+i\partial_y, \partial\bar{\partial}=\Delta$. This
 is the system $GKMMN-II$ and reduction   $B_2-II$.

{\bf Theorem 2}. For the system $GKMMN-II$ in the variables
$z,\bar{z}$ the reduction to the class of self-adjoint operators $L$
with real magnetic field $-2B=2G_{\bar{z}}=F_z$ and potential $S\in
R$ is compatible with time dynamics in the variable $it,t\in R$ (V):
$$[(c_t-4c_{xy})c^{-1}]_z=8u_{xy},(u+4u_{xy})_{\bar{z}}=2/i[(u_{\bar{z}}c_z/c)-(u_{\bar{z}}c_{\bar{z}}/c)_{\bar{z}}]$$
Here we have $S=u_{\bar{z}}\in R,c=e^{2\Phi}\in R$, and system can
be written in the form (VI):
$$c_t-4c_{xy}=8a_yc=-4Im(u_z)c,
S_t+4S_{xy}=8[S\Phi_{xy}-S_x\Phi_y-S_y\Phi_x]$$ with $u=a+ib,
S=a_x-b_y,  a_y+b_x=0$.

The condition $S=0$ leads to the linear system $B_2-II$ (formula
(VII)):
$$c_t-4c_{xy}=T(x,y,t)c, \Delta T=0, T=8a_y\in R$$
Here $G=1/2(\log c)_z, B=-1/2\Delta(\log c)\in R$.

For the self-adjoint factorizable operator $L$  (i.e. $S=0$), the
Reducible Riemann Surface $\Gamma$ admits an anti-involution
$\sigma:\Gamma'\rightarrow\Gamma''$ and back, $\sigma^2=1$. {\bf The
spectrum of this operator determines the spectrum of Purely Magnetic
Nonrelativistic 2D Pauli Operator for the particles with spin 1/2.
The theory of ground states for the Algebro-Geometric Pauli
Operators is developed in [4]}. Results of the present work (without
theorem 2) can be found in our article in arXiv (see [5]).

{\bf References.}

1.S.Manakov, Uspekhi Math Nauk, 1976 v 31 n 5 pp 245-246

2.B.Dubrovin, I.Krichever, S.Novikov, Doklady AN SSSR, 1976, v 229 n
1 pp 15-18

3.A.Veselov, S.Novikov, Doklady AN SSSR, 1984, v 279 v 4 pp 784-788

4.P.Grinevich, A.Mironov, S.Novikov, arXiv: 1004.1157

5.P.Grinevich, A.Mironov, S.Novikov, arXiv: 10014300

6.B.Konopelchenko, Inverse Problems, 1988, v.4, pp 151-163.
\end{document}